\documentclass[11pt]{article}

\newfont{\tenmsb}{msbm10 scaled\magstep1}

\let\ssection=\section
\renewcommand{\section}{\setcounter{equation}{0}\ssection}

\newcommand{\half}{{\scriptstyle{\frac{1}{2}}}}
\newcommand{\cL}{{\cal L}}

\newcommand{\const}{\mathop{\rm const.}\nolimits}

\def\p{{\partial}}
\def\ve{{\vec e}}
\def\vE{{\vec E}}

\def\vp{{\vec p}}

\def\vQ{{\vec Q}}
\def\vj{{\vec\jmath}}
\def\vq{{\vec q}}

\def\vv{{\vec v}}

\def\vA{{\vec A}}
\def\vnabla{{\vec\nabla}}
\newcommand{\crit}{\mathop{\rm crit}\nolimits}
\newcommand{\Hall}{\mathop{\rm Hall}\nolimits}
\newcommand{\red}{\mathop{\rm red}\nolimits}
\newcommand{\ext}{\mathrm{ext}}
\newcommand{\tot}{\mathop{\rm tot}\nolimits}

\begin{document}

\setlength{\baselineskip}{16pt}

\title{
Exotic plasma as classical Hall Liquid
}

\author{
C.~Duval
\\
Centre de Physique Th\'eorique, CNRS\\
Luminy, Case 907\\
F-13 288 MARSEILLE Cedex 9 (France)
\\
Z.~Horv\'ath
\\
Institute for Theoretical Physics, E\"otv\"os
University\\
P\'azm\'any P. s\'et\'any 1/A\\
H-1117 BUDAPEST (Hungary)
\\
and
\\
P.~A.~Horv\'athy
\\
Laboratoire de Math\'ematiques et de Physique Th\'eorique\\
Universit\'e de Tours\\
Parc de Grandmont\\
F-37 200 TOURS (France)
}

\date{\today}

\maketitle

\begin{abstract}
A non-relativistic plasma model endowed with an ``exotic'' structure
associated with
the two-parameter central extension of the planar Galilei group is constructed.
Introducing a Chern-Simons statistical
gauge field provides us with a self-consistent system;
when the magnetic field takes a critical value determined by the
extension parameters, the
fluid becomes incompressible and moves collectively,
according to the Hall law.
\end{abstract}


\smallskip
\noindent\texttt{cond-mat/0101449}
(Revised version). To appear in {\it Int. Journ. Mod. Phys.} {\bf B}.

\goodbreak
\section{Introduction}

In a recent paper \cite{DH}, the ground states of the Fractional Quantum Hall
Effect (FQHE), represented by the ``Laughlin'' wave functions
\cite{LAUGH, QHE}, were derived by taking advantage of the two-fold
``exotic'' extension \cite{LL} of the planar Galilei group.
Our clue has been that the two extension parameters $m$ and $k$ combine with
the magnetic field into an {\it effective} mass,
\begin{equation}
m^*=m-\frac{ekB}{m}.
\end{equation}
The main result in \cite{DH} says that
for vanishing effective (rather than real \cite{DJT}) mass,
$m^*=0$, i.e., when
the magnetic field takes the (constant) critical value
\begin{equation}
B=B_{\crit}\equiv \frac{m^2}{ek},
\label{thecondition}
\end{equation}
the consistency of the equations of motion requires that
the particle move  with the Hall velocity
$
\dot{q}_{i}=v_{i}^{\Hall}\equiv
\varepsilon_{ij}{E_j}/{B},
$
with $i,j=1,2$.
Intuitively,  in uniform electric
and magnetic fields, the cyclotronic
motion of an ordinary charged particle
can, for a specific initial velocity,
degenerate to a straight line \cite{QHE}.
For an exotic particle with zero
effective mass this is the only allowed motion.
\goodbreak

\goodbreak
The generalization to $N$ particles interacting through dynamical
gauge fields \cite{CWWH} is readily seen to be inconsistent
with our vanishing effective mass condition (\ref{thecondition}),
though. Fortunately,
the ground state of the FQHE is actually described, however,
by a self-consistent,
incompressible quantum liquid, presented as a novel
state of matter~\cite{LAUGH, Stone, Wen, edge, Hallfluid}, rather than
by a single particle
moving in an external field.
Below we construct, starting with the $1$-particle model of
\cite{DH} and
following the general principles of plasma physics \cite{Plasma}, an
 ``exotic'' plasma model. When the magnetic field takes the critical
 value~(\ref{thecondition}), our plasma reduces to an incompressible
fluid which moves according to the Hall law. It can be viewed,
 hence, as the classical
counterpart of the quantum liquid in~\cite{LAUGH, Stone, Wen, edge, Hallfluid}.
Requiring that the statistical gauge field have
a Chern-Simons dynamics yields finally self-consistent solutions,
which represent the ground state of the Hall fluid.
\goodbreak

\section{Exotic particle -- gauge field system}


The planar Galilei group has long been known to admit a non-trivial
two-parameter
central extension \cite{LL}. One of the extension parameters, present in any
dimension, is conventional~: it appears in the commutator of translations
and boosts
and is interpreted as the mass, $m$.
The other, ``exotic'' one, denoted by ${k}$, only appears in two space
dimensions; it
comes from the commutator of the Galilean boosts.
Our fundamental assumption
is to view both parameters as physical.

In \cite{DH} we found that
minimal coupling to an arbitrary planar
electromagnetic field  $\vE$ and $B$ yields the equations of motion
\begin{equation}
\left\{\begin{array}{l}\displaystyle
m^*\dot{q}_i=
p_i-\frac{ek}{m}\varepsilon_{ij}E_j,
\\[4mm]
\displaystyle
\dot{p}_i=e\left(E_i+B\,\varepsilon_{ij}\dot{q}_j\right).
\end{array}\right.
\label{eqmotion}
\end{equation}

The fields  $\vE$ and $B$ here
satisfy the homogeneous Maxwell equation,
 implying that
they derive from potentials $V$ and $\vA$, respectively.
Note that $\vp$ and $m\dot{\vq}$ are different. Eliminating $\vp$ in favor of
$\dot{\vq}$ allows us to present (\ref{eqmotion}) as
\begin{equation}
m^*\ddot{q}_i
=
e
\Big(E_i+\varepsilon_{ij}\dot{q}_jB\Big)
-\frac{e k}{m}\varepsilon_{ij}\Big(\dot{q}_k\partial_kE_j
+\partial_tE_j
+\varepsilon_{jk}\dot{q}_k\big(\dot{q}_\ell\partial_\ell{}B
+\partial_t{}B\big)
\Big),
\label{modlorentz}
\end{equation}
which shows that the ``exotic'' structure results in modifying
the Lorentz force.

In \cite{DH} we analyzed our system in Souriau's  symplectic
framework \cite{SSD} (actually equivalent to ``Faddeev-Jackiw''
reduction \cite{FaJa}).
Let us explain our results using Poisson brackets.
Setting $\xi=(\vq, \vp\,)$,
the equations of motion (\ref{eqmotion}) can indeed be
written in the Hamiltonian form
\begin{equation}
    \dot{\xi}=\big\{\xi,h\big\},
    \qquad
    h=\frac{{\vp}{\,}^2}{2m}+eV(\vq,t),
    \label{hameq}
\end{equation}
the ``exotic'' Poisson bracket
 being given by \cite{DH}
\begin{equation}
\begin{array}{cc}
\big\{f,g\big\}=&\displaystyle\frac{m}{m^*}
\sum_{i=1}^2\p_{q_i}f\,\p_{p_i}g
-
\p_{q_i}g\,\p_{p_i}f
\hfill\\[4mm]
&+
\displaystyle\frac{{k}}{mm^*}\left[\p_{q_1}f\,\p_{q_2}g
-\p_{q_1}g\,\p_{q_2}f\right]
+
eB\displaystyle\frac{m}{m^*}
\left[\big(\p_{p_1}f\,\p_{p_2}g-\p_{p_1}g\,\p_{p_2}f\big)
\right].
\hfill
\end{array}
\label{exoticPB}
\end{equation}

The first term here is the conventional one;
the second one
 combines the ``exotic'' structure and
the magnetic field.
Note that the plane became consequently non-commutative~:
the coordinates satisfy
\begin{equation}
\big\{q_{1}, q_{2}\big\}= \frac{k}{mm^*}
\end{equation}
rather than commute. Let us emphasize that the non-commutativity of the
plane here  arises even in the absence of  any
gauge field, and
 {\it follows} rather directly from the assumed ``exotic'' structure.

Let us emphasize that
these formul{\ae} are valid for {\it any} planar  field.
In particular,
the ``exotic'' Poisson bracket  (\ref{exoticPB}) satisfies,
 despite the presence of the {\it a priori}
position-dependent quantity~$m^*$, the
Jacobi identity as long as the electromagnetic field satisfies the
homogeneous Maxwell equation, as it can be verified by a tedious
cal\-cu\-lation. (A quicker proof is obtained
using the associated symplectic structure.)
\goodbreak

\goodbreak
Further insight is gained by observing that, when $B$ is
constant such that $m^*\neq0$,
\begin{equation}
\left\{
\begin{array}{l}
Q_{i}=q_{i}+
\displaystyle\frac{1}{eB}\left(1-\sqrt{\frac{m^*}{m}}\right)
\varepsilon_{ij}p_{j}\hfill\\[3.6mm]
P_{i}=\displaystyle\sqrt{\frac{m^*}{m}}\,p_i
-\frac{1}{2} eB\varepsilon_{ij}Q_{j}\hfill
\end{array}
\label{cancoord}
\right.
\end{equation}
are canonical coordinates on
the $1$-particle phase space. The ``exotic'' Poisson bracket
(\ref{exoticPB}) becomes
\begin{equation}
\big\{F,G\big\}=
\sum_{i=1}^2\p_{Q_i}F\,\p_{P_i}G
-
\p_{Q_i}G\,\p_{P_i}F,
\end{equation}
so that the $4D$  volume element reads
$dQ_{1}\wedge dQ_{2}\wedge dP_{1}\wedge dP_{2}$.

In these coordinates, the canonical structure retains  hence the
standard form, while the Hamiltonian becomes, however, rather complicated.

For vanishing effective mass, $m^*=0$,
the coordinates
$
Q_{i}
$
and momenta,
$
P_{i}=-(eB/2)\varepsilon_{ij}Q_{j}
$,
are no more independent and  the Poisson bracket (\ref{exoticPB})
becomes singular.
Then symplectic reduction yields a $2$-dimensional reduced phase space with
canonical coordinates \cite{DH}
\begin{equation}
Q_{i}=q_{i}-\frac{mE_{i}}{eB_{\crit}^2}.
\label{oldQcoord}
\end{equation}
The reduced Hamiltonian and  Poisson bracket are
\begin{equation}
H\equiv H_{\red}=eV(\vQ),\hfill
\quad\hbox{and}\quad
\big\{F,G\big\}_{\red}
=
-\frac{1}{eB_{\crit}}\left(
\p_{Q_{1}}F\,\p_{Q_{2}}G-\p_{Q_{1}}G\,\p_{Q_{2}}F
\right),
\label{redhampoisson}
\end{equation}
respectively.
The new coordinates satisfy now
\begin{equation}
\big\{Q_{1}, Q_{2}\big\}_{\red}=-\frac{1}{eB_{\crit}},
\end{equation}
and  the equations of motion,
\begin{equation}
\dot{\vQ}=\{\vQ, H\}_{\red},
\end{equation}
become, by (\ref{redhampoisson}),
$
\dot{Q}_i=v^{\Hall}_i\equiv
\varepsilon_{ij}{E_j}/{B_{\crit}},
$
i.e., the Hall law. 
Note that the condition $m^*=0$ plainly requires a constant
magnetic field $B=B_{\crit}$,
whereas $\vE$ is an otherwise arbitrary curlfree
electric field.
Also observe that the reduced Hamiltonian
is just the potential expressed
in terms of the non-commuting coordinates
$Q_{i}$~: this is the so-called ``Peierls substitution''
\cite{DJT, DH}.

Note for further reference that
the  $4D$ volume element became also degenerate; the reduced volume element is
$eB_{\crit}\,dQ_{1}\wedge dQ_{2}$. Thus, while the magnetic term drops out
from the $4D$ volume element, it is the only one left after reduction.

It is worth mentioning that our reduced structure would allow us to
recover  the
infinite-dimensional symmetry of the FQHE states,
consisting of the area-preserving diffeomorphisms of the plane
\cite{Winfty, MaStone}.  These latter are in fact precisely the
 canonical transformations of the symplectic plane, i.e., the
 transformations that preserve the reduced Poisson bracket
(\ref{redhampoisson}).

To conclude this outline of the $1$-particle theory, let us
mention that the reduced model corresponds to the classical version
of the  LLL states.
Quantization allows us to recover in fact  the FQHE
ground states, represented by the ``Laughlin'' wave functions
\cite{LAUGH, QHE, DH}.

\goodbreak
\section{``Exotic'' plasma}

Adapting the general framework of \cite{Plasma} to our case, let us consider
$N$ identical ``exotic'' particles, interacting with some given abelian
gauge field;
each of them satisfies therefore the equations of motion
(\ref{eqmotion}). The volume element on the
($4N$-dimensional) $N$-particle phase space is
\begin{equation}
\prod_{a=1}^N\frac{m_{a}^*}{m}\,dp^a_{1}\wedge dp^a_{2}\wedge
dq^a_{1}\wedge dq^a_{2},
\qquad
\mbox{where}
\qquad
m_{a}^*=m-\frac{ek}{m}B_a,
\label{volumelement}
\end{equation}
and $B_a\equiv B(\vq_a)$; here the $m^*_{a}$ come from the exotic
structure.
It is worth noting that the magnetic term $eBdq_{1}\wedge dq_{2}$ does
not contribute to
the volume element, since it drops out from the square of the symplectic form.
(Similarly, when replacing the mechanical momenta, $p_{i}$,
by the canonical momenta,
$p_{i}-eA_{i}$, the gauge potentials would drop out by the same reason.)

Let us consider a distribution function $f(\vq_a,\vp_a,t)$
on phase space.
 According to Liouville's theorem, the volume element (\ref{volumelement})
 is invariant w.r.t. the classical dynamics, and $df/dt=0.$
Using the equations of motion, this means that
\begin{equation}
\frac{\p f}{\p t}+
\displaystyle\sum_{a}\frac{m}{m_{a}^{*}}\left[
\left(\frac{p_{i}^a}{m}-\frac{e{k}}{m^2}\,\varepsilon_{ij}E_{j}^a\right)
\frac{\p f}{\p q_{i}^a}
+e
\left(E_{i}^a+B^a\varepsilon_{ij}\frac{p_{j}^a}{m}\right)
\frac{\p f}{\p p_{i}^a}\right]
=0,
\label{Liouvilleeqn}
\end{equation}
where $E_{j}^a=E_{j}(\vq_a)$.
It is worth mentioning that (\ref{Liouvilleeqn}) is
indeed
\begin{equation}
\p_{t}f+\big\{f,h\big\}=0,
\end{equation}
where
$h=\sum_{a}\left(\vp_a\right)^{2}/2m+eV(\vq_a)$ is the $N$-particle
Hamiltonian, and $\big\{\cdot,\cdot\big\}$ denotes the $N$-particle
Poisson bracket
$\big\{f,g\big\}=\sum_{a}\big\{f,g\big\}_a$.

Let us first assume that the effective mass does not vanish,
$m^*\equiv m^{*}_{1}\neq0$.
Following the ``regressive (BBGKY) method''
\cite{Plasma}, we integrate  over the last $(N-1)$-particle
phase space and define the $1$-particle distribution $\phi$ as
\begin{equation}
\phi=\frac{m}{m^*}\displaystyle\int f
\prod_{a=2}^N{\frac{m^*_a}{m}\,d\vec{p}_a\,d\vec{q}_a}.
\label{phidef}
\end{equation}
 Integrating over the last $(N-1)$ particles and suppressing the
 particle label $a=1$ allows us to infer
(see App. 8.1 of Ref. \cite{Plasma}),
the  novel (Boltzmann) transport equation
\begin{equation}
\frac{\p\phi}{\p t}+
\frac{1}{m^*}
\left(p_{i}-\frac{e{k}}{m}\,\varepsilon_{ij}E_{j}\right)
\frac{\p\phi}{\p q_{i}}
+\frac{m}{m^*}
e\left(E_{i}+\frac{B}{m}\,\varepsilon_{ij}p_{j}\right)
\frac{\p\phi}{\p p_{i}}
+
\frac{ek}{m m^*}\,\dot{B}\,\phi
=0,
\label{exoticplasma}
\end{equation}
where
$
\dot{B}=\partial_{t}B+\dot{\vq}\cdot\vnabla{B}
$
is the material (or convective) derivative.
In (\ref{exoticplasma}) a complicated expression called the
collision
integral, representing the two and
more particle interactions \cite{Plasma},
has been put to zero. This is justified since the
collisions of our particles can indeed be neglected,
owing to their
infinitely short-range $\delta$-type interactions
(see (\ref{CSpot}) below).


 Our final step is to consider the mean matter density, the mean
velocity, and the mean current by averaging over the last remaining momentum
$\vp\equiv \vp_1$, namely
\begin{equation}
\varrho=\int\!\phi\, d\vp,
\qquad
\vv=\frac{1}{\varrho\,}\int\dot{\vq}\,\phi\, d\vp,
\qquad
\vj=\varrho\,\vv.
\label{meancharge}
\end{equation}
Then (\ref{exoticplasma}) yields
the hydrodynamical equations
\begin{equation}
\displaystyle\p_{t}\varrho
+\vnabla\cdot(\varrho\vv\,)=0,
\label{hydro1}
\end{equation}
\begin{equation}
\varrho\big(\displaystyle\p_{t}\vv
+
\vv\cdot\vnabla\vv\big)
=\vec{f}-\vnabla\cdot\sigma,
\label{hydro2}
\end{equation}
where $\vec{f}=\displaystyle\int\ddot{\vq}\,\phi\,d\vp$ is
the mean value of the force on the r.h.s. of (\ref{modlorentz}),
and $\sigma=(\sigma_{ij})$ is the kinetic stress tensor~\cite{Plasma}.
Owing to the infinitely short-range forces,
the inter-particle pressure can be neglected, and
$\sigma$ retains, hence, the form
\begin{equation}
\sigma_{ij}=
\displaystyle\int(\dot{q}_{i}-v_{i})(\dot{q}_{j}-v_{j})\,\phi\,d\vp.
\label{stresstensor}
\end{equation}

This statement follows from the general discussion
in \cite{Plasma}, Chap. 9. Reassuringly, it can also be shown directly~:
firstly, the continuity equation
comes from the transport equation, using Stokes' theorem
and the homogeneous Maxwell equation
$\partial_{t}B+\vnabla\times\vE=0$.

The exotic structure only enters the force. This latter is indeed found,
using (\ref{modlorentz}), to be
\begin{equation}
f_i=\frac{e\varrho}{m^*}
\left[
E_i+\varepsilon_{ij}v_jB
-\frac{k}{m}\varepsilon_{ij}\Big(\dot{E}_j
+\varepsilon_{jk}v_k\dot{B}
\Big)
\right]
+\frac{ek}{mm^*}\sigma_{ij}\partial_j{}B,
\label{force}
\end{equation}
where $\dot{E}_j=v_k\partial_kE_j+\partial_tE_j$ and
$\dot{B}=v_k\partial_k{}B+\partial_t{}B$.
Then, the Euler equation (\ref{hydro2}) follows from the modified
force law (\ref{modlorentz}) [or from (\ref{force})]
 by a tedious calculation.

A look at the $N$--particle transport equation (\ref{exoticplasma})
shows now that, in the limit $m^*\to0$,
the consistency requires that the coefficients of  $1/m^*$ vanish~:
\begin{eqnarray}
    p_{i}-e\frac{k}{m}\,\varepsilon_{ij}E_{j}=0,
    \qquad
E_{i}+\frac{B}{m}\,\varepsilon_{ij}p_{j}=0,
\label{hallp}
\end{eqnarray}
yielding the  Hall law. 
This same condition can also be obtained from  the
hydrodynamical equations. A tedious calculation yields in fact that
the Hall law is necessary for the consistency of (\ref{hydro2}).

Further insight is gained by rewriting, for constant $B$ and nonvanishing
$m^{*}$,
(\ref{exoticplasma}) in terms of the twisted position coordinates $\vQ$
and the original momenta, $\vp$, as
\begin{equation}
\p_{t}\phi+\dot{Q}_{i}\frac{\p\phi}{\p Q_{i}}+
\varepsilon_{ij}\frac{eB}{m^*}
\left(p_{j}-m\varepsilon_{jk}\frac{E_{k}}{B}\right)\frac{\p\phi}{\p{p}_{i}}=
0,
\label{exoticplasmabis}
\end{equation}
where
\begin{equation}
\dot{Q}_{i}=\varepsilon_{ij}\frac{E_{j}}{B}
+
\frac{1}{\sqrt{mm^*}}\Big(p_{i}-m\varepsilon_{ij}\frac{E_{j}}{B}\Big).
\end{equation}
It follows, as in (\ref{hallp}), that in the limit
$m^*\to0$ the vector $\vp$ and hence
also $\dot{\vQ}$ have to satisfy the
Hall constraint, namely
\begin{equation}
p_i=m\dot{Q}_i=m\varepsilon_{ij}\frac{E_j}{B}.
\label{Hallconstraint}
\end{equation}

\goodbreak
Next, for $m^*\to0$,
the $4N$-dimensional phase space ``shrinks'' to a $2N$-dimensional
reduced phase space, and the very definition (\ref{phidef}) of the $1$-particle
distribution $\phi$ becomes meaningless.
The reduced quantities can {\it not} be obtained by setting simply
$m^*=0$~: the
physical quantities may not behave continuously as $m^*\to0$ \cite{DJT}.
The whole construction of Section 2 has to be repeated therefore once
again, using the
reduced structures. Let us hence consider a distribution function
$F\equiv{}F_{\red}(\vQ,t)$ on reduced phase space. Then Liouville's equation
(\ref{Liouvilleeqn}) is replaced, using the reduced
Hamiltonian structure (\ref{redhampoisson}),
by
\begin{equation}
\p_tF+\{F,H\}_{\red}=\p_tF-\frac{\vE\times\vnabla F}{B_{\crit}}=0.
\label{redhameq}
\end{equation}
The reduced $1$-particle distribution on $2D$ phase space,
\begin{equation}
\Phi\equiv\Phi_{\red}(\vQ)=
\int\! F\,\prod_{a=2}^N{eB_{\crit}\,dQ_1^adQ_2^a}
\label{redPhi}
\end{equation}
satisfies therefore
\begin{eqnarray}
\p_t\Phi+\frac{\varepsilon_{ij}E_j}{B_{\crit}}\,\frac{\p\Phi}{\p Q_{i}}=0,
\label{redPhieq}
\end{eqnarray}
which replaces, for $m^*=0$, the fundamental equation
(\ref{exoticplasmabis})
by fixing the velocity $\dot{\vQ}$
and putting the  term proportional to $1/m^*$ to zero.

The mean matter density $\varrho$ is in fact $\Phi$ in (\ref{redPhi}).
Since all particles are frozen in a collective Hall motion, integrating out the
momenta  in (\ref{meancharge}) amounts to restricting the currents to the $2D$
surface in $4D$ phase space, defined by the Hall constraint
(\ref{Hallconstraint}). In fact,
lifting $\Phi$ to the original phase space as
$\phi(t,\vQ,\vp)=\Phi(t,\vQ)\,\delta\big(\vp-m\vv_{\Hall}\big)$,
the mean charge and velocity in (\ref{meancharge}) become
\begin{equation}
    \varrho=\Phi,
    \qquad
    \vv=\vv_{\Hall}.
\end{equation}
The density hence satisfies
\begin{equation}
\p_{t}{\varrho}
+\varepsilon_{ij}\frac{E_{j}}{B_{\crit}}\,
\frac{\p\varrho}{\p Q_{i}}=0,
\label{redcont}
\end{equation}
which is clearly is a Hamiltonian equation w.r.t. the reduced
structure.

Note that our equation (\ref{redcont}) is, indeed, consistent with the
continuity equation (\ref{hydro1}) for the current
$
\vj=~\vv\varrho,
$
because $\vnabla\cdot\vv=(1/B)\vnabla\times\vE=0$
thanks to the homogeneous Maxwell equation $\p_{t}B+\vnabla\times\vE=0$
with $B=B_{\crit}$.
The fluid is therefore {\it incompressible}.
It thus admits,
as any incompressible fluid in the plane,
the infinite dimensional symmetry of area-preserving
diffeomorphisms\cite{Arnold},
found above for a single particle.

Let us observe that no
Euler equation analogous to (\ref{hydro2}) is obtained here,
since the mean velocity is entirely determined by the Hall law.
(This is somehow analogous to the drop in the phase space dimension.)

In conclusion, our results obtained so far say that for vanishing effective
mass the consistency
requires that the fluid move according to
the Hall law, with the velocity determined by the gauge field
$B=B_{\crit}$ and $\vE$ felt by the particle,
{\it whatever is the origin} of this latter.
\goodbreak

\goodbreak
\section{Coupled Chern-Simons -- matter system, and the Hall states}

 The dynamics of the gauge field has not been
 specified so far.  To this end,
let us consider a  coupled matter-gauge field system
 described by an action
$S\!=\!\displaystyle\int\!\cL_{\rm matter}+\cL_{\rm gauge-field}$.
To be consistent with the fundamental galilean symmetry of our approach,
we choose for $\cL_{\rm gauge-field}$  the
 Chern-Simons Lagrangian  \cite{DJTe}, also
including external magnetic and electric fields,
$\vE_{\ext}$ and~$B_{\ext}$, respectively.
Generalizing the matter Lagrangian of \cite{CWWH}, we add $N$
``exotic'' terms (actually equivalent to
the second-order Lagrangian of Lukierski et al. \cite{LSZ}, and
is consistent with
the symplectic form used in \cite{DH}). Thus,
we describe our $N$  identical exotic particles with mass
$m$, exotic structure ${k}$  and charge $e$,
minimally coupled to a Chern-Simons
gauge field $(A_{\mu})=(A_t,\vA)$, by the action
\begin{equation}
\begin{array}{cc}
S=&\displaystyle\sum_{a=1}^N\displaystyle\int\!
 \big(\vp_{a}-e\vA\big)\cdot d\vq_{a}
-\left[\frac{(\vp_{a})^{2}}{2m}
+eA_{t}\right]dt
+
\displaystyle\frac{{k}}{2m^2}\varepsilon_{ij}p_{i}^{a}dp_{j}^{a}
\hfill\\
\\
&+\;\kappa\displaystyle\int\left\{\displaystyle\frac{1}{2}
\varepsilon_{\alpha\beta\gamma}F_{\alpha\beta}A_{\gamma}
-B_{\ext}A_{t}-\vA\times\vE_{\ext}
\right\}d\vq\,dt,
\hfill\\
\end{array}
\label{claction}
\end{equation}
where $\kappa$ is a new (Chern-Simons) coupling constant.
 Variation w.r.t. the particle coordinates yields $N$ ``exotic''
matter equations (\ref{eqmotion}),
and variation w.r.t. the gauge field yields the
Chern-Simons field equations
\begin{equation}
{\kappa}\,B=-e\varrho_{\tot},
\qquad
{\kappa}\,\varepsilon_{ij}E_{j}=-e{\jmath}_{i}^{\tot},
\label{CSeqn}
\end{equation}
the external fields being hidden
here  in the total  density and current,
 $(j_\mu^{\tot})=(\varrho_{\tot},\vec\jmath_{\tot})$,
 defined as
$j_\mu^{\tot}=\delta S/\delta A_{\mu}$, {\it viz}.
\begin{equation}
\begin{array}{lll}
e\varrho_{\tot}
&=e\varrho-\kappa B_{\ext}
&=\displaystyle\sum_{a}e\delta(\vq-\vq_{a})-\kappa B_{\ext},
\\[3.6mm]
e\jmath_{i}^{\tot}
&=e\jmath_{i}-\kappa\varepsilon_{ij}E_{j}^{\ext}
&=\displaystyle\sum_{a}ev^{a}_i\delta(\vq-\vq_{a})
-\kappa\varepsilon_{ij}E_{j}^{\ext}.
\end{array}
\label{partcurrents}
\end{equation}

Decomposing the total fields into the sum of
the external and ``statistical'' quantities $(b,\ve\,)$,
$B=b+B_{\ext}$ and
$\vE=\ve+\vE_{\ext}$ respectively, we see
that the gauge-field part of
(\ref{claction}) is actually equivalent to
$
\half\int
\varepsilon_{\alpha\beta\gamma}\,f_{\alpha\beta}\,a_{\gamma}\,
d\vq\,dt.
$
Thus, while the particle
feels the total gauge field, the statistical field
itself
obeys the Chern-Simons dynamics
with the particle current as source. From~(\ref{CSeqn})
we infer in fact
\begin{equation}
{\kappa}b=-e\varrho,
\qquad
{\kappa}\varepsilon_{ij}e_{j}=-e{\jmath}_i.
\label{statCSeqn}
\end{equation}

The $\delta$-functions in the particle current
represent point-like vortices in the effective-field theory
approach \cite{ZHK},
and correspond to Laughlin's quasiparticles \cite{QHE}.
It is now clear
that, owing precisely to these vortices,
$\varrho$  (and hence $B$) can never be a constant.
Thus, while the coupled
Chern-Simons gauge field system associated
with (\ref{claction}) may admit (even interesting)
solutions, our reduction trick,
which would require $B=\const$, can not work.
Hence the necessity to ``smear out'' these point-like singularities,
and use, instead, the continuum model
constructed in the previous Section.
But, before doing this, let us remember
  that the gauge fields can be eliminated by
 solving the Chern-Simons equations \cite{CWWH},
\begin{equation}
a_i(\vq,t)=
\frac{e}{2\pi\kappa}\int\!\varepsilon_{ij}
\frac{q_j-y_j}{\vert\vq-\vec{y}\,\vert^2}\varrho(\vec{y},t)\,d\vec{y},
\qquad
a_t(\vq,t)=\frac{e}{2\pi\kappa}\int\!
\frac{(\vq-\vec{y})\times\vj(\vec{y},t)}{\vert\vq-\vec{y}\,\vert^2}\,d\vec{y}.
\label{CSpot}
\end{equation}
It follows that
the inter-particle forces are infinitely short-range,
as anticipated when deriving our plasma model.
\goodbreak

Let us now  turn to considering continuum matter with Chern-Simons coupling,
i.e.,
replace the $N$-particle system with our ``exotic'' plasma.
Owing to the well-known difficulties encountered in the  action formulation of
fluid dynamics \cite{JACKMONT}, we do not insist on a variational
approach, and work, instead, directly with the equations of motion.
Thus, we {\it posit} the Chern-Simons equations (\ref{statCSeqn}),
 coupled to the fluid dynamical equations
(\ref{hydro1}-\ref{hydro2}),
for $m^*\neq0$, and (\ref{redcont}) for $m^*=0$, respectively.

We now observe that these equations are  consistent with
collective--motion {\it Ansatz}
\begin{equation}
\vj_{\tot}=\vv\,\varrho_{\tot},
\label{collmotansatz}
\end{equation}
{\it provided} the velocity is
$\vv=\vv_{\Hall}$.
Note that the ``external sources''
$\varrho_{\ext}=-(\kappa/e)B_{\ext}$ and
$\jmath^{\ext}_{i}=-(\kappa/e)\varepsilon_{ij}{E_j}^{\ext}$
can also be viewed \cite{Manton} as
background charge/current densities, which also satisfy the Hall
law.

The general  coupled
system (\ref{hydro1}-\ref{hydro2})-(\ref{CSeqn}), will be studied elsewhere.

Here our point is that when we restrict ourselves to
the case $m^*=0$, i.e.,  when the magnetic field felt by the
fluid takes the critical value $B=B_{\crit}$,
the collective-motion Ansatz~(\ref{collmotansatz})
becomes {\it mandatory} for the reduced system, yielding
$E_j/B_{\crit}= E_j^{\ext}/B_{\ext}$. The
velocity  determined by the fields felt by the
particle is, hence,  also given
 by the external field alone,
\begin{equation}
	v_{i}=\varepsilon_{ij}\frac{E_j}{B_{\crit}}=
	\varepsilon_{ij}\frac{E_j^{\ext}}{B_{\ext}}
	\equiv v_{i}^{\Hall}.
\label{GoodHallLaw}
\end{equation}

As seen above,
the total flow is incompressible; when the external field $B_{\ext}$
is also uniform, the matter density $\varrho$ becomes also
constant,  and the matter flow is also incompressible.
\goodbreak
In the critical case the  Chern-Simons equations require hence
\begin{equation}
\varrho=\frac{\kappa}{e}(B_{\ext}-B_{\crit}),
\qquad
\vj=\varrho\,\vv_{\Hall},
\label{plascurrents}
\end{equation}
with $\vv$ obeying the Hall law (\ref{GoodHallLaw}).
Equation. (\ref{plascurrents})
represents hence the ground state of the Hall fluid.
If the electric field is, in addition,  divergence-free,
$\vnabla\cdot\vE=0$, (e.g., if the external fields are uniform),
then the flow becomes also  irrotational.
When $B_{\ext}=B_{\crit}$, the particle density
vanishes, $\varrho=0$. Hence
there is no statistical field, merely
 a uniform background charge $\varrho_{\ext}$,
 which moves according to the Hall law.
When the external field
is moved out from the critical value, excitations are created~:
the quantity
$\varrho\!=\!\varrho_{\tot}-\varrho_{\ext}$ describes in fact the deviation
from the background density.
If the external fields are uniform,
so is $\varrho$ : the excitations condensate into collective modes.
The sign of $\varrho$
is positive or negative depending on that of $(B_{\ext}-B_{\crit})$,
corresponding to quasiparticles and
quasiholes, respectively \cite{LAUGH, QHE, Hallfluid}.

According to the Chern-Simons equations (\ref{CSeqn})
$\varrho$ and $\vj$ are the sources of the statistical field,
whose r\^ole is to maintain the the total magnetic field at the critical value
$B_{\crit}$ (and to create an electric field such that the ``external Hall
law''
(\ref{GoodHallLaw}) holds).

\section{Conclusion}

In this paper we have derived
an exact, self-consistent solution, (\ref{plascurrents}),
of the coupled exotic
matter--Chern-Simons gauge field system.
Our solution  is associated  with
 vanishing effective mass, $m^*=0$, i.e., with the  magnetic field
 taking the critical value  $B_{\crit}=m^2/ek$.
 Intuitively, we want to view the limit $m^{*}\to0$ as
 ``condensation into the collective ground state'';
 in other words, a kind of
``phase transition'' into this strongly correlated
 ``novel  state of matter''
\cite{QHE, LAUGH, Stone, edge, Hallfluid} we identify with the FQH
ground state.
We are not in the position to prove, within our classical context, that
$m^*=0$ would be mandatory. Our investigations imply, however,  that
when this condition holds, then the Hall motions are the only
consistent ones.
It is worth to be mentioned, however, that our reduced
fluid dynamical equation (\ref{redcont})
has been proposed  to describe the chiral bosons of
edge currents  \cite{Wen}, which indicates that
our vanishing effective mass
condition $m^*=0$  may be physically relevant.

Quantum fluids, non-commutative structures, and
Chern-Simons theory have already been considered in this
context by many authors; see, e.g.,
\cite{LAUGH, Stone, Wen, edge, Hallfluid, noncomm, NCCS}.
The approach closest to ours would be that of Stone \cite{Stone},
who derives
 the Euler equations of fluid dynamics using the
 ``Madelung'' transcription of the effective
 ``Landau-Ginzburg'' theory \cite{ZHK}.
 Our approach here is, however, rather different~:
 it is entirely based
 on the classical model  associated with the
 ``exotic'' structure of the planar Galilei group.
 In particular, the non-commutativity of the
plane follows from this structure unlike in other approaches \cite{NCCS}.

It has been noticed before \cite{DJT} that
a reduced model leading to the ground states of the QHE
can be obtained by letting the ordinary mass, $m$, go to zero.
 Our
``exo\-tic'' model allows us, however, to avoid taking such an
unphysical limit~: our vanishing effective mass condition,
$m^*=0$, only requires to fine-tune the magnetic field
to its critical value determined by the
parameters $m$ and $k$ (assumed here physical).
Note, however, that in our approach the ``good'' coordinates
are the ``twisted coordinates''
$Q_{i}$, and not the physical coordinates $q_{i}$.

Our fluid model is derived straightforwardly from the modified
force law (\ref{modlorentz}), following the general principles of plasma
physics~\cite{Plasma}.
Galilean invariance would actually allow to
add a magnetic (but no electric) Maxwell term \cite{Manton},
which would contribute a term
$\varepsilon_{ij}\partial_jB$ in the
second Chern-Simons equation (\ref{CSeqn}). This
would not change our conclusions, though, since the new term
 would drop out since $B=B_{\crit}=\const$

Our results are consistent with some of the essential properties of
Hall fluids and constitute therefore a strong argument in support of
the physical relevance of the ``exotic'' Galilean structure.
 Let us insist that, despite its physical dimension
 $[k]=[\hbar/c^2]$,
 our ``exotic parameter'' is a {\it classical} object.
 (Remember that the phase space symplectic form
 $dp_{i}\wedge dq_{i}$ has also dimension $[\hbar]$.)
It is, just like anyonic spin, unquantized.
As pointed out by Jackiw and Nair \cite{JaNa},
$k$ can be viewed
as a kind of non-relativistic  ``shadow'' of relativistic spin.

At last, our derivation shows indeed some similarity with that of
Martinez and Stone \cite{MaStone},
who obtain the Hall law as the first
approximation to their   second-quantized equation of motion.
 This latter
corresponds in fact to the quantized version of
our classical equation of motion (\ref{redhameq}), their Moyal bracket being
the quantum deformation of our classical Poisson bracket.
Similarly, the classical symmetry of area-preserving
transformations, $w_{\infty}$ is replaced, under quantization, by its
quantum version $W_{\infty}$ \cite{Winfty, MaStone}.
Our theory might hence be viewed as a classical counterpart
of that of Martinez and Stone.
\goodbreak

\goodbreak

\section{Acknowledgements}

We are indebted to M. Stone for correspondence and advice,
and to R.~Jackiw and P.~Forg\'acs for discussions.
Z.~H. acknowledges
the {\it Laboratoire de Math\'emathiques et de Physique Th\'eorique}
of Tours University for hospitality.
\goodbreak


\end{document}